\documentclass{aastex}
\usepackage{spr-astr-addons}
\usepackage{url}
\usepackage{float}

\usepackage{hyperref}
\usepackage[usenames,dvipsnames]{xcolor}

\newcommand{\emaila}{\href{mailto:arrincon@uc.cl}{arrincon@uc.cl}}

\begin{document}

\title{On non-coplanar Hohmann Transfer using angles as parameters}

\slugcomment{Not to appear in Nonlearned J., 45.}
\shorttitle{Short article title}
\shortauthors{Rinc\'on et al.}
\author{\'Angel Rinc\'on\altaffilmark{1,2}}
\altaffiltext{1}{Departamento de F\'isica, P. Universidad Cat\'olica de Chile, Casilla 306, Santiago, Chile. E-mail: \emaila}

\and 

\author{Patricio Rojo\altaffilmark{2}}
\altaffiltext{2}{Departamento de Astronom\'ia, Universidad de Chile, Casilla 36-D, Santiago, Chile.}

\and

\author{Elvis Lacruz\altaffilmark{3}}
\altaffiltext{3}{Centro de Investigaciones de Astronom\'ia, A.P. 264, C.P. 5101, M\'erida, Venezuela.}
 
\and

\author{Gabriel Abell\'an\altaffilmark{4}}
\altaffiltext{4}{Departamento de F\'isica, Universidad Central de Venezuela, AP 47270, Caracas 1041-A, Venezuela.}

\and

\author{Sttiwuer D\'iaz\altaffilmark{5}}
\altaffiltext{5}{Grupo de Informaci\'on y Comunicaci\'on Cu\'antica, Departamento de F\'isica, Universidad Sim\'on Bolivar, Sartenejas, Edo. Miranda 89000, Venezuela.}

\begin{abstract}
We study a more complex case of Hohmann orbital transfer of a satellite by considering non-coplanar and elliptical orbits, instead of planar and circular orbits. We use as parameter the angle between the initial and transference planes that minimizes the energy, and therefore the fuel of a satellite, through the application of two non-tangential impulses for all possible cases. We found an analytical expression that minimizes the energy for each configuration. Some reasonable physical constraints are used: we apply impulses at perigee or apogee of the orbit, we consider the duration of the impulse to be short compared to the duration of the trip, we take the nodal line of three orbits to be coincident and the three semimajor axes to lie in the same plane. 
\noindent
We study the only four possible cases but assuming non-coplanar elliptic orbits. In addition, we validate our method through a numerical solution obtained by using some of the actual orbital elements of Sputnik I and Vanguard I satellites. 
\noindent
For these orbits, we found that the most fuel-efficient transfer is obtained by applying the initial impulse at apocenter and keeping the transfer orbit aligned with the initial orbit.
\end{abstract}


\section{\label{sec:Intro} Introduction}
In 1925, Hohmann studied the transfer between coplanar circular orbits and found that the minimum fuel transfer in a Newtonian gravitational field occurs when two impulses are applied producing an elliptic transfer orbit which is tangent to both of the terminal circular orbits. A first impulse is used to set the vehicle into the elliptic transfer orbit, while a second impulse leads to a circular orbit at the final radius (\cite{Hohmann_1960,Prussing_1991}). Many researchers have made contributions to the improvement and understanding of this type of orbital transfers. More recently, Hohmann transfer has been generalized from the original idea to more general cases: \cite{Broucke_1994} considered $N$-impulse transfers between any two coplanar orbits ($\forall \, N \leqslant 4$) and for the two-impulse maneuver developed optimality conditions that lead to a non-linear system of three equations and three unknowns, whereas \cite{Arlulkar_2012} discussed Hohmann transfer between two circular orbits but including a dynamical approach Lambert solution (i.e., considering the transfer time of the orbit to change from one point to another). Reference \cite{Mabsout_2009} addressed the optimization of the orbital Hohmann transfer considering only the coplanar case using as optimization parameter the eccentricity of the transfer orbit. On the other hand, different techniques of standard optimization had been used for minimizing a cost function. For example, one studied the isoperimetric problem of finding the extremal transfers for the given characteristic velocity for the orbits \citep{Kirpichnikov_2003}.
\noindent Some of the authors of this paper \citep{Lacruz_2010} have calculated the solution for the generalized non-coplanar Hohmann transfer only for the first configuration (there exist four configurations that minimize the energy according to \cite{Kamel_1999}).
\noindent In this paper we consider elliptic orbits and $N = 2$ impulse transfers \citep{Broucke_1994}; taking a split between initial and transference planes ($i_{it} \neq 0^{\circ}$), this improves and generalizes the work of \cite{Mabsout_2009} and differs from \cite{Kirpichnikov_2003} since we will not consider the launch time of spacecraft. We consider non-coplanar orbits, which improve the solution of \cite{Arlulkar_2012}.
\noindent It is relevant to discuss the role of orbital transfer in astronomy and engineering: the orbital transfers are required for a standard space mission. Generalized  coplanar Hohmann transfer had been used to model a space vehicle traveling in elliptic orbits of the Earth and Jupiter around the Sun \citep{Kamel_2011} and shows the importance of this kind of study in astronomy and the planetary sciences.
\noindent The standard (non-perturbed) transfer between orbits is treated using Kepler problem theory, i.e. considering Keplerian orbits, because these are non-perturbed solutions of the two body problem as a first approximation to a typical orbital motion. We choose to work with the standard set of inertial orbital elements: 
 $\mathcal{O} := \{a, e, i, \Omega, \omega, \tau\}$, 
where $a$ is the \textit{semimajor axis}, $e$ is the \textit{eccentricity}, $i$ is the \textit{inclination}, 
$\Omega$ is the longitude of the \textit{ascending node} and $\omega$ is the \textit{argument of periapsis}. 
Finally, the sixth parameter is the \textit{epoch} $\tau$ indicating the time at which the orbiter passes through periapsis.
\noindent Using the solution of the Kepler problem and some constraints in orbital elements, we develop an extension of the original Hohmann model considering elliptical and non-coplanar orbits. We investigate orbital changes between elliptical orbits using non-tangential impulses which are applied at periapsis and apoapsis of the orbit in order to obtain minimum cost of fuel. In this paper we find several minimum solutions and we determine which case \citep{Kamel_1999} is optimal from an energetic point of view.
\noindent The present paper is organized as follows. In Sect. \ref{sec:Problem} we discuss the problem and the physical constraints used to solve it. In Sect. \ref{sec:M&R} we use a standard optimization technique for each configuration and get a polynomial function, whose solution (once we found its inverse) gives the angle between the initial orbit and transfer orbit. In Sect. \ref{sec:NE&D} we consider some numerical values in order to use our solution in a particular example whereas in Sect. \ref{sec:Dis} we discuss briefly the solution and consequences. Finally, in Sect. \ref{sec:conclu} we present some relevant conclusions of this paper.
\section{\label{sec:Problem} Problem to solve: A general approach}
In order to visualize all possible trajectories of an orbiter we consider the initial, transfer, and final orbits, with the same nodal line (see Fig. \ref{Fig1}). The initial orbit is where first impulse is applied and has a set of orbital parameters given by
$\mathcal{O}(a, e, i, \Omega, \omega, \tau)_{i}$. The second orbit, the transfer orbit, is described by a set of parameters $\mathcal{O}(a, e, i, \Omega, \omega, \tau)_{t}$; this orbit is where the second impulse will be applied. The arrival orbit has orbital parameters $\mathcal{O}(a,e,i,\Omega, \omega, \tau)_{f}$. We want to emphasize that each orbit lies in a plane and between any two planes it is possible to define an inclination angle. This is an important point since we want to find the minimum value given by the minimization of cost function (usually the cost function is defined as the sum of impulses per unit mass), taking one of the orbital elements as parameter. This will be commented on in Sect. \ref{sec:M&R}. 
\subsection{Model Constraints}
\begin{itemize}
\item The radius of the initial impulse is lower than the radius of the final impulse.
\item The initial, $\mathcal{O}_i$, and final, $\mathcal{O}_f$, orbits form an inclination 
angle $i_{if} \in (0^{\circ} , 180^{\circ} )$, between their orbital 
planes $\mathcal{P_O}_i$ and $\mathcal{P_O}_f$, respectively.
\item When we apply the first impulse $\Delta v_1= ||\Delta \vec{v}_1||$, where 
$\Delta \vec{v}_1$ is the vector of the first maneuver, it produces an inclination angle $i_{it} \neq 0^{\circ} $, 
between the $\mathcal{P_O}_i$  and
$\mathcal{P_O}_t$ plane of the transfer orbit, $\mathcal{O}_t$.
\item When we apply the second impulse $\Delta v_2= ||\Delta \vec{v}_2||$, where 
$\Delta \vec{v}_2$ is the vector of the second maneuver, it produces an inclination angle $i_{tf}$, between the $\mathcal{P_O}_t$,  and $\mathcal{P_O}_f$. 
\item The apsides of the three orbits are collineal with the nodal line of the three orbits.
\item The primary focus, $F_p$, is common in the three orbits and coincides with the origin of inertial reference frame.
\end{itemize}
 \begin{figure}[ht]
  \centering
  \begin{picture}(230,180)
\put(0,0){\includegraphics[scale=.3]{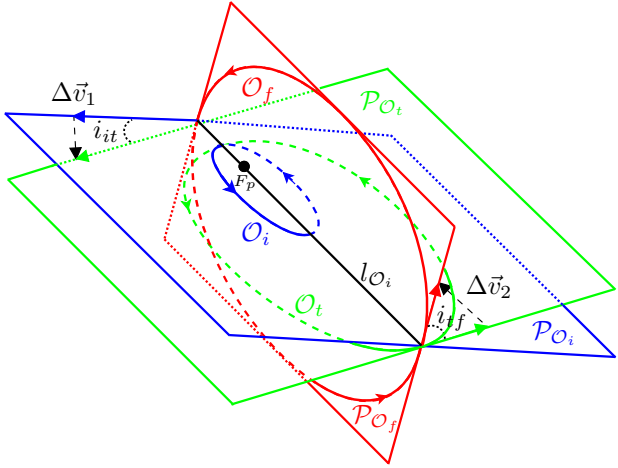}}    
    \put(90,140){\color{red}$\mathcal{O}_{f}$}
    \put(132,16){\color{red}$\mathcal{P}_{\mathcal{O}_{f}}$}
    \put(90,85){\color{blue}$\mathcal{O}_{i}$}
    \put(200,48){\color{blue}$\mathcal{P}_{\mathcal{O}_{i}}$}
    \put(110,58){\color{green}$\mathcal{O}_{t}$}
    \put(135,135){\color{green}$\mathcal{P}_{\mathcal{O}_{t}}$}
    \put(18,138){$\Delta \vec v_{1}$}
    \put(175,65){$\Delta \vec v_{2}$}
    \put(34,124){$i_{it}$}
    \put(164,54){$i_{tf}$}
    \put(135,70){$l_{\mathcal{O}_{i}}$}
    \put(87,106){{\tiny $F_{p}$}}
    \end{picture}    
       \caption{$\mathcal{P}_{\mathcal{O}_i}$ is the plane (blue color) where the initial orbit is located, $\mathcal{P}_{\mathcal{O}_t}$ is the plane (green color) that contains the transfer orbit. $\mathcal{P}_{\mathcal{O}_f}$ is the plane (red color) that contains the final orbit. The arrow directions indicate the motion of the satellite in each orbit. $l_{\mathcal{O}_{i}}$ is the so-called line of apsides for the initial orbit. Note that it is not required for the initial orbit to be interior to the projection final orbit we just take $r_{f} \gg r_{i}$ as a particular case.}
  \label{Fig1}
\end{figure}  
\noindent In Fig. \ref{Fig1} we show the three angles $i_{if}$, $i_{it}$ and $i_{tf}$. Now, the angle $i_{if}$ is fixed and defined (because we know in advance the angle between initial and arrival orbits), so we will choose one of the angles $i_{it}$ and $i_{tf}$ as the parameter to minimize. In order to do the minimization, we need to define a function that allows us to calculate the angles previously mentioned. This is the necessary energy for the 
orbital maneuvers. We call this function $\mathcal{F}$ the \textit{cost function}; it may depend on all orbital parameters.
\subsection{Mathematical aspects}
\cite{Hohmann_1960} found the transfer of minimum cost between two circular orbits using an elliptical transfer orbit. In this work we choose the inclination $i_{it}$ as parameter. However, it is possible to choose any orbital parameter in order to find a minimum of cost function and get the best possible trajectory \citep{Abad_2012}.
\noindent We need to write down the cost function in terms of the two impulses. In order to compute the impulses, we need to get the norm of each one and relate them with the orbital parameters. This is given by the {\it Vis Viva} equation \citep{Montenbruck_2001},
 \begin{align}
  \epsilon = -\frac{\mu}{2a} &=  \frac{1}{2}v^2 - \frac{\mu}{r}, 
  \label{visviva}
 \end{align}
\noindent where $\epsilon$ is the orbital energy, $\mu = GM_{\oplus}$ is a constant 
(with $M_{\oplus}$ Earth mass), $r$ is the relative distance between the two bodies, and $v=||\vec{v}||$, where $\vec{v}$ is the required velocity vector.
\noindent Since the two impulses will be applied, one at the perigee and the other at the apogee, we need to obtain the velocities in perigee and apogee for each orbit. Using the formulation for a general conic section we get, in polar coordinates,
 \begin{subequations}
\begin{align}
 r_{a} &= (1+e)a,     \label{ra}
 \\ 
 r_{p} &= (1-e)a.     \label{rp}
 \end{align}
\end{subequations}
Considering Eqs. (\ref{ra}), (\ref{rp}), and (\ref{visviva}) it is possible to obtain the norm of velocity in perigee and apogee which induces a simple solution, 
\begin{subequations}
\begin{align}
 v_{a} &=\sqrt{\frac{\mu}{a} \Bigl(\frac{1-e}{1+e} \Bigl)}. 
 \label{va}
\\
 v_{p} &= \sqrt{\frac{\mu}{a} \Bigl(\frac{1+e}{1-e} \Bigl)}. 
\label{vp} 
\end{align}
\end{subequations}
\noindent Using (\ref{va}), (\ref{vp}) and the standard impulse definition we get 
the two impulses applied at perigee and apogee, and we define the cost function $\mathcal{F}$ as the sum of this two maneuvers. In the following section we will show this procedure in detail.
\section{\label{sec:M&R} Method and results}
We define the cost function $\mathcal{F}$ as
\begin{align}
 \mathcal{F} & \equiv   \parallel \Delta \vec{v}_1 \parallel + \parallel \Delta \vec{v}_2 \parallel \label{F},
\end{align}
\noindent where these impulses (per unit mass) are non-tangen\-tial, applied at apside extreme lines, that is perigee and apogee, respectively (see Fig. \ref{Fig2}). The impulsive maneuver vectors are given by
\begin{align}
\Delta \vec{v}_1 &\equiv \vec{u}_i - \vec{v}_i, \\
\Delta \vec{v}_2 &\equiv \vec{v}_f - \vec{u}_f,
\end{align}
\noindent where the vectors $\vec{v}_i$ and $\vec{u}_i$ refer to the initial velocities of the initial and transfer orbits, respectively. In the same way, $\vec{v}_f$  and $\vec{u}_f$ are the final velocities of the final and transfer orbits, respectively. The first {impulse, $\Delta v_1=||\Delta \vec{v}_1||$, is applied in the initial orbit and the second impulse is applied in the transfer orbit, necessary to switch to the final orbit. When the first impulse is applied, the initial and transfer orbits coincide; that happens just in this point, as is seen in Fig. \ref{Fig2}, for different configurations.
\begin{figure*}[ht]
\begin{center}
\begin{picture}(280,280)
\put(-90,150){\includegraphics[scale=.40]{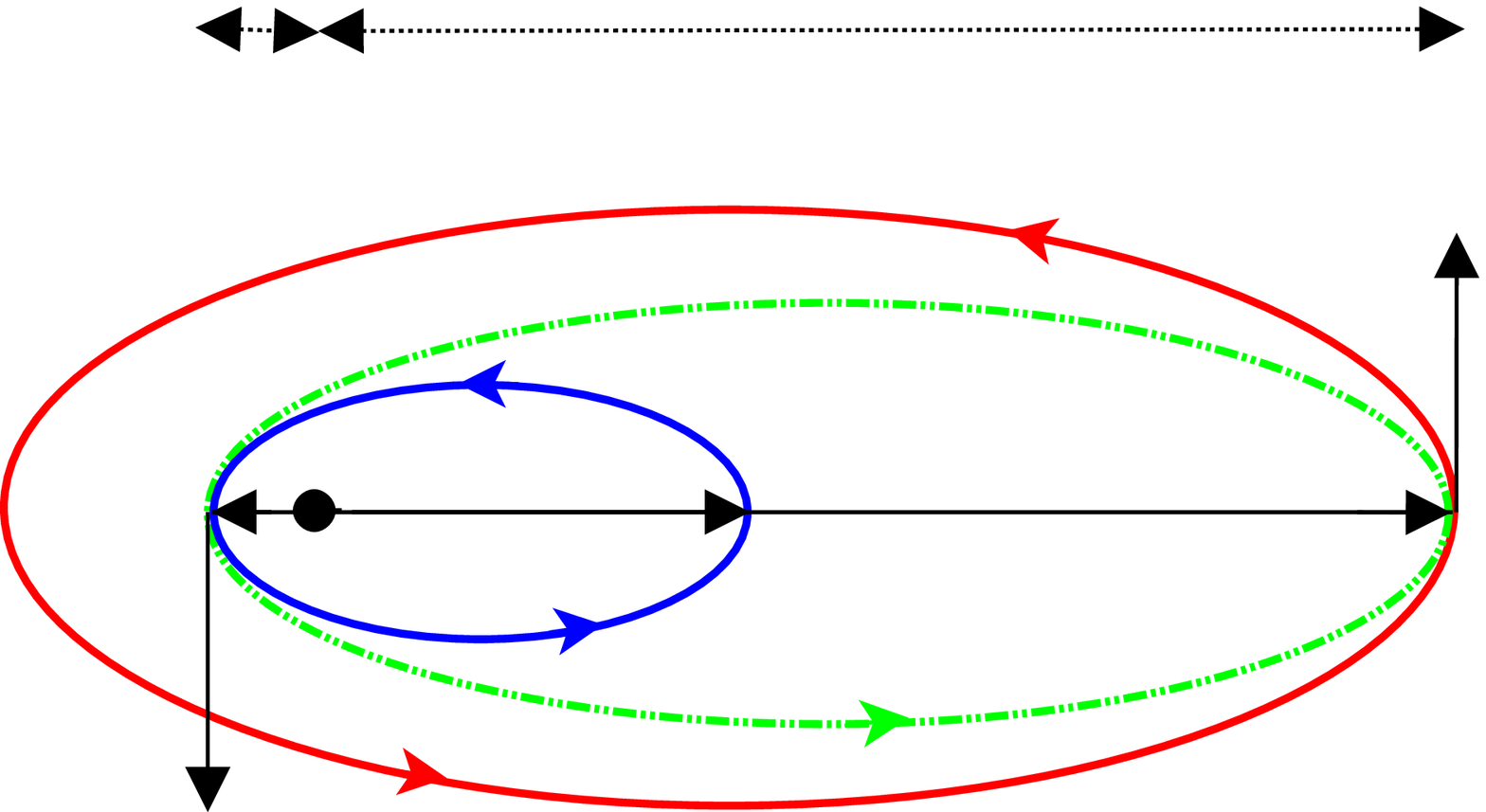}}
    \put(-54,272){$r_{i}$}
    \put(32,272){$r_{f}$}
    \put(-60,235){\color{red}$\mathcal{O}_{f}$}
    \put(13,205){\color{blue}$\mathcal{O}_{i}$}
    \put(10,168){\color{green}$\mathcal{O}_{t}$}
    \put(40,186){\small $r_{a_{\color{green}\mathcal{O}_{t}}} = r_{a_{\color{red}\mathcal{O}_{f}}} $}
    \put(-53,188){\scriptsize $r_{p_{\color{blue}\mathcal{O}_{i}}}$}
    \put(-40,198){\small $F_{p}$}
    \put(-80,158){$\Delta \vec{v}_{1}$}
    \put(127,220){$\Delta \vec{v}_{2}$}
    \put(-10,135){$r_{p_{\mathcal{O}_i}} = r_{p_{\mathcal{O}_t}} \equiv r_{i}$}
    \put(-12,122){$r_{a_{\mathcal{O}_t}} = r_{a_{\mathcal{O}_f}} \equiv r_{f}$}
\put(160,150){\includegraphics[scale=.40]{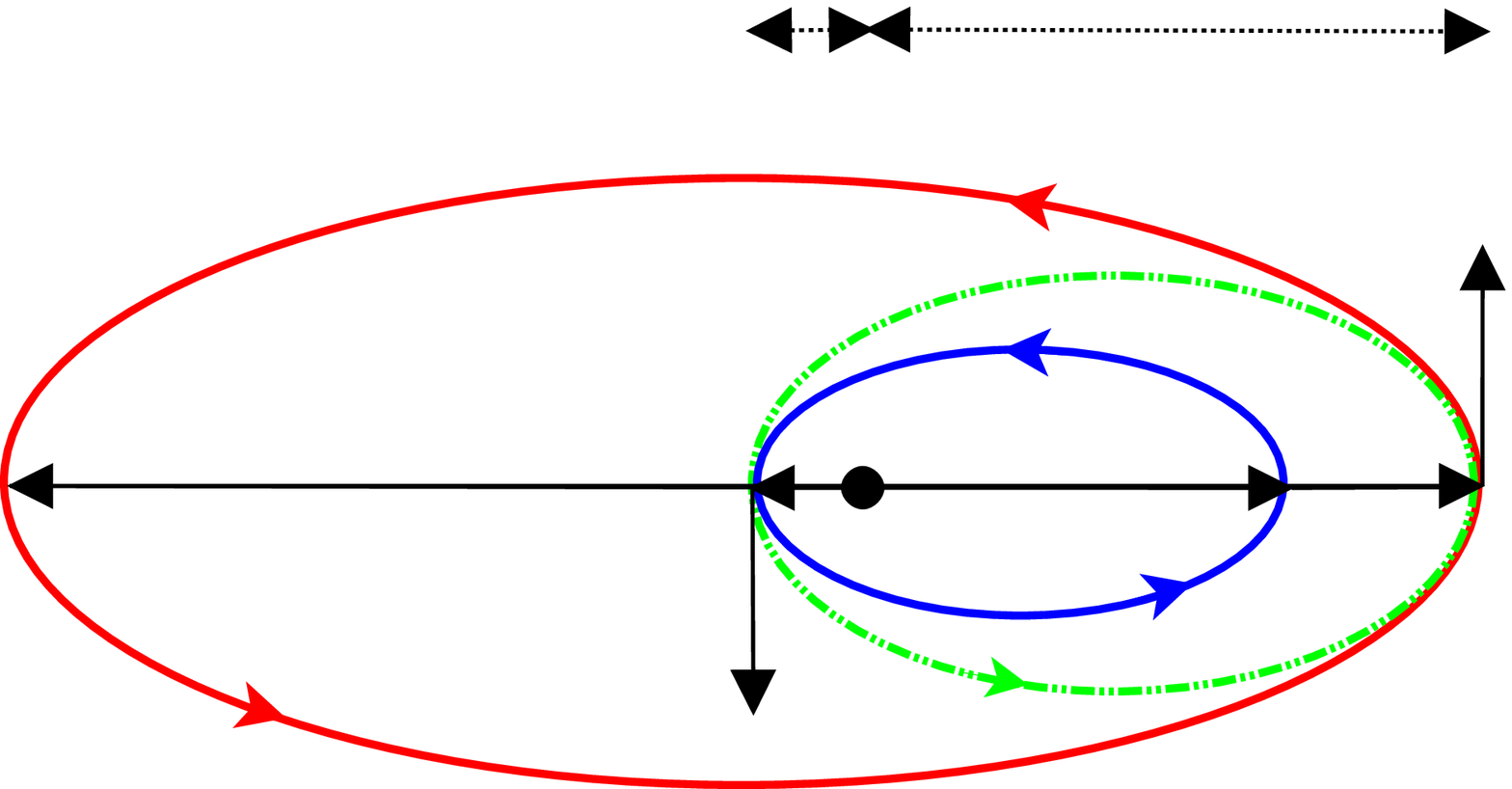}}
  \put(275,272){$r_{i}$}
    \put(325,272){$r_{f}$}
    \put(190,235){\color{red}$\mathcal{O}_{f}$}
    \put(340,205){\color{blue}$\mathcal{O}_{i}$}
    \put(275,160){\color{green}$\mathcal{O}_{t}$}
    \put(190,186){\small $r_{a_{\color{red}\mathcal{O}_{f}}}$}
    \put(275,188){\scriptsize $r_{p_{\color{blue}\mathcal{O}_{i}}}$}
    \put(305,186){\small $r_{a_{\color{green}\mathcal{O}_{t}}}$}
    \put(350,186){\small $r_{p_{\color{red}\mathcal{O}_{f}}}$}
    \put(282,198){\small $F_{p}$}
    \put(248,170){$\Delta \vec{v}_{1}$}
    \put(375,215){$\Delta \vec{v}_{2}$}
    \put(236,135){$r_{p_{\mathcal{O}_i}} = r_{p_{\mathcal{O}_t}} \equiv r_{i}$}
    \put(235,122){$r_{p_{\mathcal{O}_f}} = r_{a_{\mathcal{O}_t}} \equiv r_{f}$}
\put(-90,-30){\includegraphics[scale=.32]{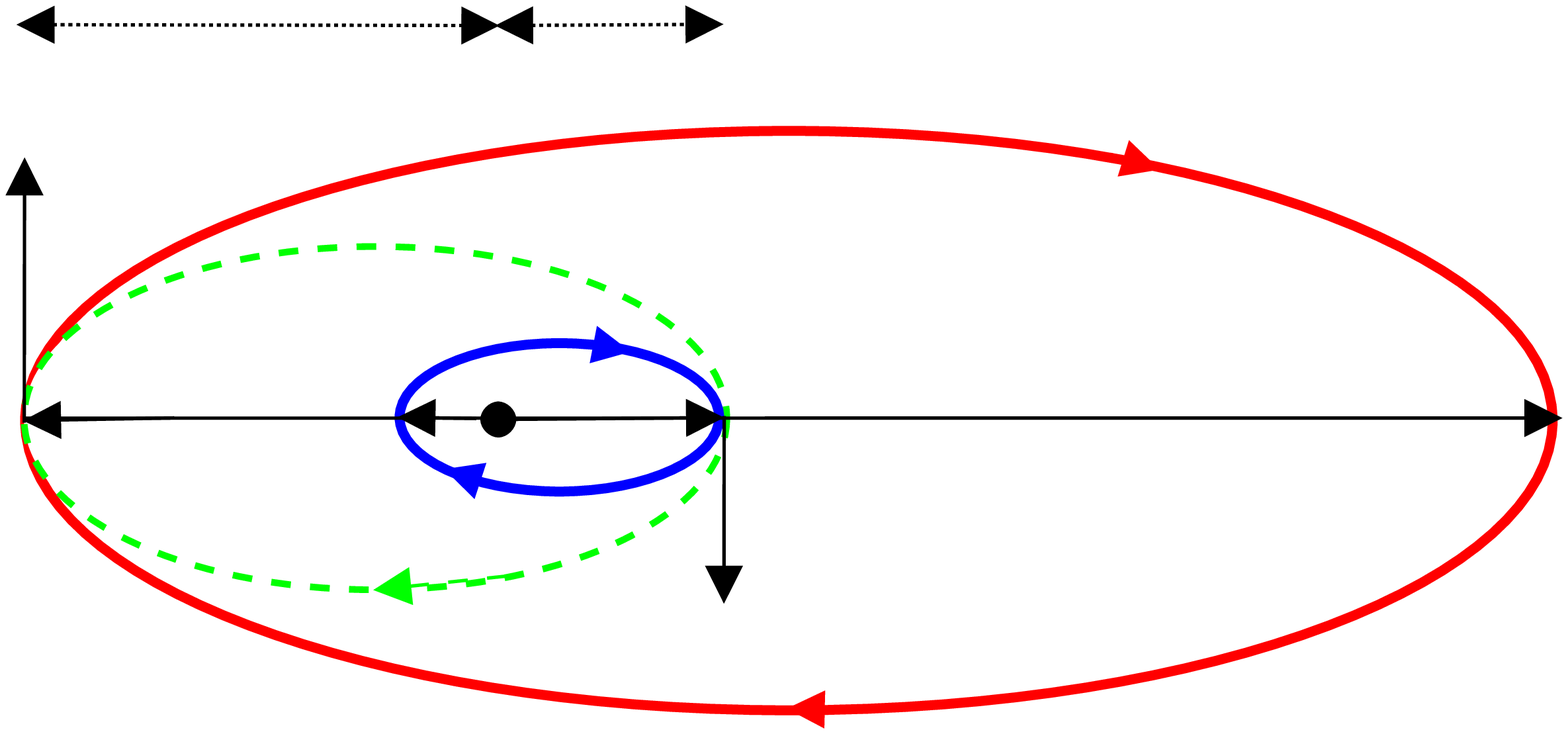}}
    \put(-60,80){$r_{f}$}
    \put(-12,80){$r_{i}$}
    \put(-60,50){\color{red}$\mathcal{O}_{f}$}
    \put(-35,25){\color{blue}$\mathcal{O}_{i}$}
    \put(-15,-15){\color{green}$\mathcal{O}_{t}$}
    \put(-17,9){\scriptsize $r_{a_{\color{blue}\mathcal{O}_{i}}}$}
    \put(50,8){\small $r_{a_{\color{red}\mathcal{O}_f}}$}
    \put(-55,8){\scriptsize $r_{p_{\color{red}\mathcal{O}_f}}$}
    \put(-17,16){\small $F_{p}$}
    \put(12,-2){$\Delta \vec{v}_{1}$}
    \put(-83,45){$\Delta \vec{v}_{2}$}
    \put(-11,-42){$r_{a_{\mathcal{O}_i}} = r_{a_{\mathcal{O}_t}} \equiv r_{i}$}
    \put(-12,-53){$r_{p_{\mathcal{O}_f}} = r_{p_{\mathcal{O}_t}} \equiv r_{f}$}
\put(160,-30){\includegraphics[scale=.40]{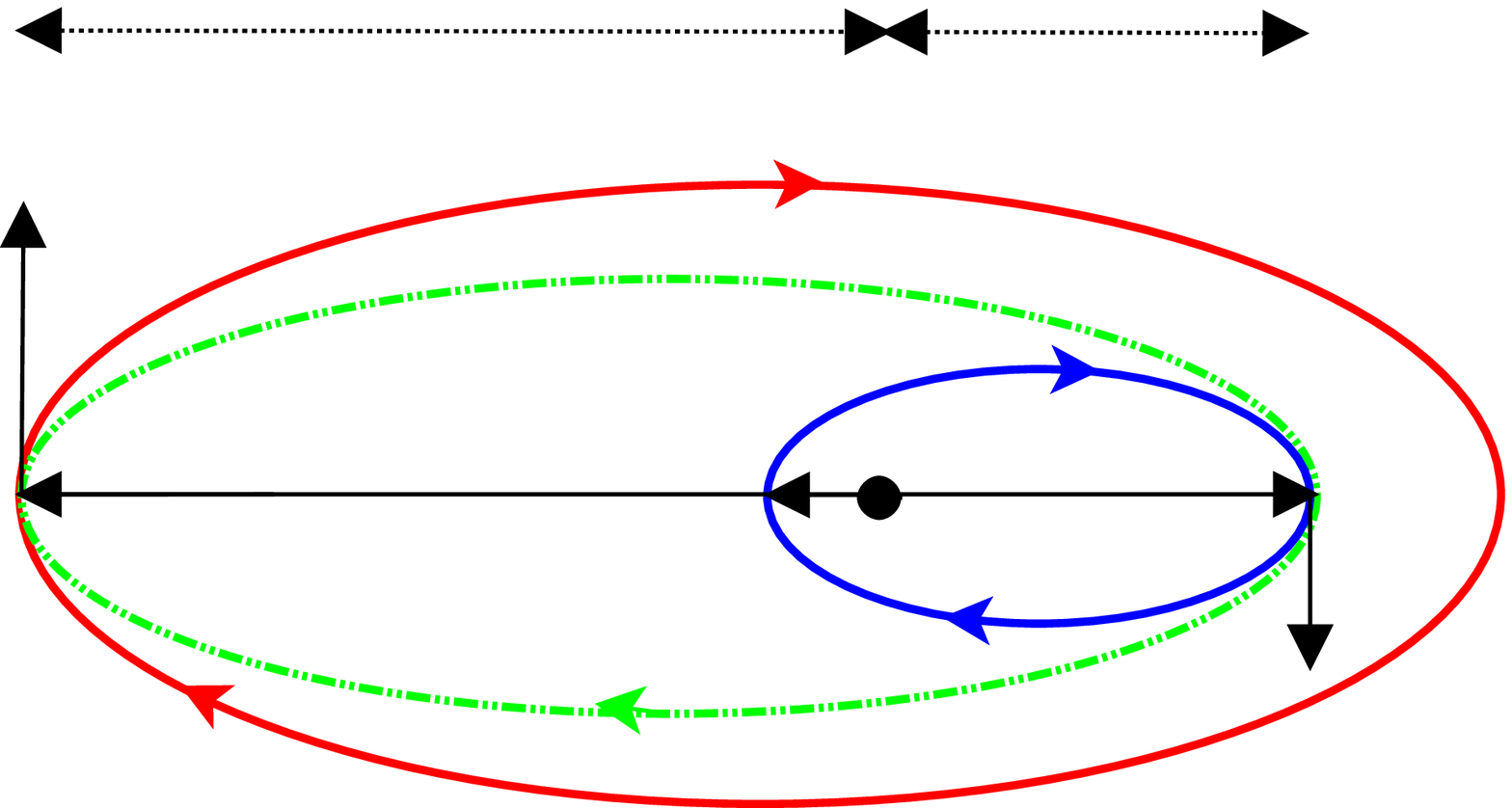}}
  \put(220,92){$r_{f}$}
    \put(310,92){$r_{i}$}
    \put(190,58){\color{red}$\mathcal{O}_{f}$}
    \put(342,24){\color{blue}$\mathcal{O}_{i}$}
    \put(275,-11){\color{green}$\mathcal{O}_{t}$}
    \put(195,7){\small $r_{a_{\color{red}\mathcal{O}_{f}}} = r_{a_{\color{green}\mathcal{O}_{t}}}$}
    \put(283,5){\small $r_{a_{\color{blue}\mathcal{O}_{i}}} = r_{p_{\color{green}\mathcal{O}_{t}}}$}
    \put(282,20){\small $F_{p}$}
    \put(347,0){$\Delta \vec{v}_{1}$}
    \put(168,46){$\Delta \vec{v}_{2}$}
    \put(236,-42){$r_{a_{\mathcal{O}_i}} = r_{p_{\mathcal{O}_t}} \equiv r_{i}$}
    \put(235,-53){$r_{a_{\mathcal{O}_f}} = r_{a_{\mathcal{O}_t}} \equiv r_{f}$}
\end{picture}
\end{center}
\vspace{2cm}
\caption{Configurations considering two non-tangential impulses applied in perigee and apogee orbit (only four cases). We show constraints relative to this for each case. The figure in the upper left panel corresponds to our first configuration, the upper right figure corresponds to the second. The lower left and right figures are the third and fourth configurations, respectively. These figures are a plane projection of the three-dimensional problem \citep{Lacruz_2010}.}
\label{Fig2}
\end{figure*}
\noindent For the first  configuration (upper left panel of Fig. \ref{Fig2}) the initial impulse $\Delta \vec{v}_1$ is located at the perigee of the transfer elliptic orbit $r_{p_{\mathcal{O}_{t}}}$ and this coincides with the initial orbit, whereas the final impulse $\Delta \vec{v}_2$ is located at the apogee of the transfer elliptic orbit $r_{a_{\mathcal{O}_{t}}}$ and this coincides with the final orbit. Note that the other three configurations are easily obtained using the corresponding impulse maneuvers according to Fig. \ref{Fig2}.
\noindent Thus, a link between the parameters of the orbits is established. A similar situation occurs between the transfer and arrival orbits when the second impulse is applied. We found an expression for the impulse maneuvers in terms of the velocity vectors for each orbit and angles between orbital planes; their norms are
\begin{align}
 \parallel\Delta \vec{v}_1 \parallel&  = \sqrt{u_i^2 + v_i^2 - 2 u_i v_i \cos(i_{it})}           \label{Dv1},\\
 \parallel\Delta \vec{v}_2 \parallel&  = \sqrt{v_f^2 + u_f^2 - 2 u_f v_f \cos(i_{if} - i_{it})},  \label{Dv2}
\end{align}
\noindent where $u_{i}=||\vec{u}_{i}||$, $v_{i}=||\vec{v}_{i}||$, $u_{f}=||\vec{u}_{f}||$, and $v_{f}=||\vec{v}_{f}||$. Notice that we used the fact $i_{if} \equiv i_{it} + i_{tf}$.
\noindent To find the minimum value for the $\mathcal{F}$ function it is necessary to calculate the derivative with respect to one of the orbital parameters or another free parameter. As we already said, we can choose any parameter we want. The variation respect to $i_{it}$ has been little studied in the literature, so we decided to work this case.
\noindent In order to find the minimum value with respect to $i_{it}$ for the function $\mathcal{F}$,  it is required that
\begin{subequations}
\begin{align}
 \frac{\partial \mathcal{F}}{\partial i_{it}} &= 0    \label{F1}, 
\\
 \frac{\partial^2 \mathcal{F}}{\partial i_{it}^2} &>0 \label{F2}.
 \end{align}
\end{subequations}
\noindent Putting (\ref{Dv1}) and (\ref{Dv2}) into (\ref{F1}) and rearranging terms in a convenient way
\begin{align}
 \frac{u_i v_i \sin(i_{it})}{v_f u_f \sin(i_{if} - i_{it})} = \sqrt{\frac{\mathcal{K}_{i} - 2 u_i v_i \cos(i_{it})}{\mathcal{K}_{f} - 2 u_f v_f \cos(i_{if} - i_{it})}}, \label{op}
\end{align}
\noindent with $\mathcal{K}_{i}$ and $\mathcal{K}_{f}$ being the kinetic energies before and after applying the impulses, per mass unit, respectively. To facilitate the algebra we make the following definitions:
\begin{subequations}
\begin{align}
 \sigma &= \cos(i_{it}),                                                                                   \label{s1}\\
 1-\sigma^2 &= \sin^2(i_{it})                                                                             \label{s2},\\
 \cos(i_{if} - i_{it}) &= \mathcal{A}_{1}\sigma + \mathcal{A}_{2}\sqrt{ 1-\sigma^2}                             \label{s3},\\
 \sin^2(i_{if} - i_{it}) &=\mathcal{A}_{1}^{2} + \mathcal{A}_{3}\sigma^2 + \mathcal{A}_{4}\sigma\sqrt{1-\sigma^2},
 \label{s4}
\end{align}
\end{subequations}
\noindent where the $\mathcal{A}_{\ell}$ set, with $\ell = 1, 2, 3, 4$, is given by
\begin{align}
\mathcal{A}_{\ell} = \{\cos(i_{if}), \sin(i_{if}), -\cos(2i_{if}), -\sin(2i_{if})\}.
\end{align}

\noindent Substituting Eqs. (\ref{s1}), (\ref{s2}), (\ref{s3}) and (\ref{s4}), into (\ref{op}) and after some calculations we obtain a polynomial of the form \cite{Lacruz_2010}
\begin{align}
 \sum_{j=0}^{j=3} b_j \sigma^j = \sqrt{1 - \sigma^2}\sum_{j=4}^{j=6} b_j \sigma^{j-4} \label{sum1},
\end{align}
\noindent where the constant $b_j$ are
\begin{subequations}
\begin{align}
 b_0 &=  u_i^2 v_i^2 (v_f^2 + u_f^2)               -   v_f^2 u_f^2 (u_i^2 + v_i^2)\cos^2(i_{if}), \\ 
 b_1 &= 2v_f^2 u_f^2 u_i v_i \cos^2(i_{if})        -  2u_i^2 v_i^2 v_f u_f \cos(i_{if})         , \\
 b_2 &=  v_f^2 u_f^2 (u_i^2 + v_i^2)\cos(2i_{if})  -   u_i^2 v_i^2 (v_f^2 + u_f^2)              , \\
 b_3 &= 2u_i^2 v_i^2 v_f u_f \cos(i_{if})          -  2v_f^2 u_f^2 u_i v_i \cos(2i_{if})        , \\
 b_4 &= 2u_i^2 v_i^2 v_f u_f \sin(i_{if})                                                       , \\
 b_5 &= -v_f^2 u_f^2 (u_i^2 + v_i^2)\sin(2i_{if})                                               , \\
 b_6 &= 2v_f^2 u_f^2 u_i v_i \sin(2i_{if})         -  2u_i^2 v_i^2 v_f u_f \sin(i_{if}).         
\end{align}
\end{subequations}
\noindent Squaring the Eq. (\ref{sum1})  and regrouping terms, we obtain the following sixth-degree polynomial function
\begin{align}
 \mathcal{P}(\sigma) &= \sum_{j=0}^{j=6}c_j\sigma^j = 0 \label{sum2},
\end{align}
\noindent where the $c_j$ are
\begin{subequations}
\begin{align}
c_0 &=b_0^2 - b_4^2,\\
c_1 &=2(b_0b_1 - b_4b_5),\\
c_2 &=2(b_0b_2 - b_4b_6) + b_1^2 +b_4^2 - b_5^2,\\
c_3 &=2(b_0b_3 - b_5b_6 + b_1b_2 + b_4b_5),\\ 
c_4 &=2(b_1b_3 + b_4b_6) + b_2^2 +b_5^2 - b_6^2,\\
c_5 &=2(b_2b_3 + b_5b_6),\\
c_6 &=b_3^2 + b_6^2,
\end{align}
\end{subequations}
\noindent which are represented in an implicit form in terms of the velocities and angles, respectively. 
\noindent Now we need to obtain the roots of Eq. (\ref{sum2}), since they will provide the angle that minimizes the cost function. This procedure is similar in each case, but some important differences appear due to the transfer parameters eccentricity and semimajor axis \citep{Kamel_2011}. In order to keep this discussion in general terms, we need to define both parameters appropriately:
\begin{align}
a_{t}^{(n)} & = \frac{1}{2}\Bigl((1+k^{(n)}_{f}e_{f})a_{f} + (1+k_{i}^{(n)}e_{i})a_{i}\Bigl), \\
e_{t}^{(n)} & = \frac{(1+k^{(n)}_{f}e_{f})a_{f} -(1+k_{i}^{(n)}e_{i}) a_{i}}{(1+k^{(n)}_{f}e_{f})a_{f} + (1+k_{i}^{(n)}e_{i})a_{i}},
\end{align}
\noindent where ${k^{(n)}_{f}}$ and ${k^{(n)}_{i}}$ are parameters that change depending of each case (see Table \ref{tab:1} for specific values). In the same way, the initial and final velocities are different depending on the configurations. All the possible cases are shown in Table \ref{tab:2}, according to Fig. \ref{Fig2}. 
\begin{table}[t]
\centering
\caption{Selective parameter ${k^{(n)}_{f}}$ and ${k^{(n)}_{i}}$ for each case.}
\label{tab:1}       
\begin{tabular}{@{}ccccc}
\tableline
\tableline
                & Case 1 & Case 2 & Case 3 & Case 4 \\
\tableline
${k^{(n)}_{i}}$ & -1 & -1 & +1 & +1\\
${k^{(n)}_{f}}$ & +1 & -1 & -1 & +1\\
\tableline
\end{tabular}
\end{table}
\begin{table}[t]
\caption{Initial and final velocity vectors for each case, where $\vec{v}_{p_{ \mathcal{O}_i}}, \  \vec{v}_{p_{ \mathcal{O}_t}}$ and $\vec{v}_{p_{ \mathcal{O}_f}}$ are the velocity vectors in the perigee in initial, transfer and final orbits, respectively.  $\vec{v}_{a_{ \mathcal{O}_i}}, \  \vec{v}_{a_{ \mathcal{O}_t}}$ and $\vec{v}_{a_{ \mathcal{O}_f}}$ are the velocity vectors in the perigee in initial, transfer and final orbits, respectively. }
\label{tab:2}
\centering  
\begin{tabular}{@{}ccccc}
\tableline
\tableline
       & Case 1 & Case 2 & Case 3 & Case 4\\
\tableline
  $\vec{v}_{i}$ & $\vec{v}_{p_{ \mathcal{O}_i}}$ & $\vec{v}_{p_{ \mathcal{O}_i}}$ & $\vec{v}_{a_{ \mathcal{O}_i}}$ & $\vec{v}_{a_{ \mathcal{O}_i}}$\\ 
  $\vec{v}_{f}$   & $\vec{v}_{a_{ \mathcal{O}_f}}$ & $\vec{v}_{p_{ \mathcal{O}_f}}$ & $\vec{v}_{p_{ \mathcal{O}_f}}$ & $\vec{v}_{a_{ \mathcal{O}_f}}$ \\

 $\vec{u}_{i}$   & $\vec{v}_{p_{ \mathcal{O}_t}}$ & $\vec{v}_{p_{ \mathcal{O}_t}}$ & $\vec{v}_{a_{ \mathcal{O}_t}}$ & $\vec{v}_{p_{ \mathcal{O}_t}}$ \\

 $\vec{u}_{f}$   &$\vec{v}_{a_{ \mathcal{O}_t}}$ & $\vec{v}_{a_{ \mathcal{O}_t}}$ & $\vec{v}_{p_{ \mathcal{O}_t}}$ & $\vec{v}_{a_{ \mathcal{O}_t}}$ \\    
\tableline
\end{tabular}
\end{table}
%
\section{\label{sec:NE&D} Numerical example}
In order to illustrate the solution of this problem (i.e. find the minimum value of the cost function with respect to the $i_{it}$ variable), we get the cost function for each case using some known reference values. First of all, we need to get two impulse maneuvers to produce Hohmann transfer with orbital plane change. For this purpose in general six constants are required: $\mu_{\oplus} \equiv G M_{\oplus} = 3.98 \times 10^{11}$ km$^3$ s$^{-2}$, the initial-final angle $i_{if} = \pi/2$ rad, and the set $\{e_{i}, a_{i}, e_{f}, a_{f}\}$ where we have taken $e_{i} = 0.052 $, $a_{i} = 6948$ km corresponding to the Sputnik I satellite \citep{Sputnik_I}, and $e_{f} = 0.190 $, $a_{f} = 8682.5$ km corresponding to the Vanguard I satellite \citep{Vanguard_I}. 
\noindent Using previous parameters, our solutions give different roots where one of this is the minimum global in the cost function between all possible roots (six in the most general case). The roots  are in Table \ref{tab:3} displayed case by case. 
\begin{table}[!hbpt]
\small
\caption{Inclination angle obtained by two different techniques: using our sixth-degree polynomial function $\langle i_{it}\rangle_{p}$ 
and by the numerical solution $\langle i_{it}\rangle_{e}$. 
$ \Delta \langle i_{it}\rangle$ indicates absolute error in radians. The angles obtained using our procedure and the numerical solution are given in radians.}\label{tab:3}
\begin{tabular}{@{}ccccc@{}}
\tableline
\tableline
 & Case 1 &Case 2 & Case 3 & Case 4 \\
\tableline
$\langle i_{it}\rangle_{p}$  & 0.0577968 & 0.0269421 & 0.0260117  & 0.0505942\\
$\langle i_{it}\rangle_{e}$    	    & 0.0577970 & 0.0269422 & 0.0260117           & 0.0505981 \\
$ \Delta \langle i_{it}\rangle$    & 0.0000002 & 0.0000001 & 0 $\mathcal{O}(7)$  & 0.0000039\\
\tableline
\end{tabular}
\end{table}
For instance, in the Case 1 $\langle i_{it}\rangle_{p} = 0.0577968 $ rad and $\langle i_{it}\rangle_{e} = 0.0577970 $ rad, equivalently $3.31152417$ degree and $3.3115127$ degree, respectively. Therefore, the absolute error is $\Delta \langle i_{it}\rangle =|\langle i_{it}\rangle_{p}-\langle i_{it}\rangle_{e}|= 10^{-5}$ degree.
\noindent In Table \ref{tab:3}, the term $\mathcal{O}(7)$ means the exact value and the numerical values are in agreement to the seventh order of a polynomial expansion of the correct result.
\section{\label{sec:Dis} Discussion}
\noindent Since all cases share the same initial and final orbital elements, the efficiency between them can be compared. In this sense, we note that the cases 1 and 4 (such as the cases 2 and 3) can be directly compared, since the cost function looks similar. It can be understood due to cases 1-4 and 2-3 presenting some symmetry degree (reflection). 
In spite of it, the cost functions of case 1 and 4 (and case 2 and 3) are different because of the particular start and arrival points in the orbit (Fig. \ref{Fig2}). 
\begin{figure}[H]
  \centering
	\includegraphics[scale=.36]{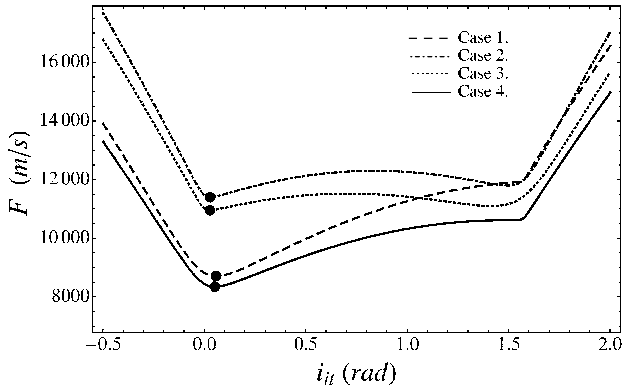}
 \caption{Cost functions (by case) $\mathcal{F}$ 
   put together as a function of the initial-transfer angle $i_{it}$. 
   Black points are the minimum values of the cost function for each case. The fourth case is the most economical case relative to the other three cases.}
   \label{Fig3}
\end{figure}
\noindent We find that two cases are the ones most efficient, the first case has the initial impulse at pericenter, with the corresponding final impulse at apocenter whereas the fourth case has the initial impulse at apocenter, with the corresponding final impulse at apocenter. On the other hand, cases 2 and 3 are the ones worst with respect to the energy used to change between orbits. 
\noindent By revisiting the cost function plot we note that the more economical configurations arrive at the apocenter whereas the more expensive cases arrive at the pericenter (Fig. \ref{Fig3}). Furthermore, as we need to apply an extra impulse to stop the vehicle from the transfer orbit to final orbit (case 2 and 3), a difference between the cost functions is established.

\section{\label{sec:conclu}Conclusions}
In this paper we obtain the analytical minimum cost function of an orbital transference between two non-coplanar elliptical orbits considering as free parameter the inclination between the initial and transfer plane. We recovered the solution obtained by \cite{Lacruz_2010} for the first configuration and calculate the solution for the other three possible cases. We compared the only four possible cases \citep{Kamel_1999} of the orbital transfer considering two impulses applied in perigee and apogee, and we determine the best model for a given set of orbital elements. We compare our analytical solution using orbital data from well-known satellites.
Comparing the exact solution and our solution, we observe that the roots where the cost function is minimal are approximately equal so our result is considered valid.
In the same way, in accordance with our solution, the fourth case is the optimal possible case. Finally, we show that it is always cheaper to have the final impulse at apocenter.

\acknowledgments
AR acknowledges support by CONICYT/ALMA Astronomy Grants $\#$ 31110010 and CONICYT-Chile through Grant D-21151658, PR was supported by Fondecyt proyect $\#$ 1120299, Basal PFB06 y Anillo ACT1120. EL acknowledges support through the CIDA. The authors would especially like to thank Daniel Casanova Ortega, who agreed to review this work.

\bibliographystyle{spr-mp-nameyear-cnd}
\bibliography{biblio_1.bib}


\end{document}